\documentclass[fleqn,10pt]{wlscirep}
\usepackage{url}
\usepackage{color}
\usepackage{float}

\usepackage[utf8]{inputenc}
\usepackage[T1]{fontenc}
\usepackage{setspace}

\onehalfspace

\usepackage{setspace}

\title{Predicting the cascading dynamics in complex networks via the bimodal failure size distribution}
\author[1]{Chongxin Zhong}
\author[1]{Yanmeng Xing}
\author[1]{Ying Fan}
\author[1,$\ast$]{An Zeng}
\affil[1]{School of Systems Science, Beijing Normal University, Beijing, P.R. China}

\affil[$\ast$]{anzeng@bnu.edu.cn}

\begin{abstract}
Cascading failure as a systematic risk occurs in a wide range of real-world networks. Cascade size distribution is a basic and crucial characteristic of systemic cascade behaviors. Recent research works have revealed that the distribution of cascade sizes is a bimodal form indicating the existence of either very small cascades or large ones. In this paper, we aim to understand the properties and formation of such bimodal distribution of cascade sizes in complex networks, and further predict the final cascade size. We first find that the bimodal distribution of cascade sizes is ubiquitous in both synthetic and real networks. Moreover, the large cascade sizes distributed in the right peak of bimodal distribution are resulted from either the failure of nodes with high load at the first step of the cascade or multiple rounds of cascades triggered by the initial failure. Accordingly, we propose a hybrid load metric (HLM), which combines the load of the initial broken node and the load of failed nodes triggered by the initial failure, to predict the final size of cascading failures. Finally, we validate the effectiveness of HLM by computing the accuracy of identifying the cascades belonging to the right and left peaks of the bimodal distribution. The results show that HLM is a better predictor than commonly used network centrality metrics in both synthetic and real-world networks.\\

\end{abstract}

\begin{document}
\maketitle
% \flushbottom
\section{\label{sec:level1}Introduction}
Cascading failure is a common phenomenon in many complex systems \cite{albert2000error, haldane2011systemic, fairley2004unruly}. The vast majority of large-scale cascading failures can be traced back to a single-element failure. With the catastrophic effects\cite{motter2002cascade, pourbeik2006anatomy}, a failure (due to a random breakdown or intentional attack) of systems can trigger further failures which may lead to the whole system being dysfunctional and collapsing. The cascades not only take place in infrastructure like electrical power grids and transportation networks \cite{yang2017small,zhao2016spatio}, but also present in social and economic systems \cite{haldane2011systemic,contreras2014propagation}. Applying network science to uncover complicated dynamic behaviors emerging from these real complex systems is an important research direction \cite{boccaletti2006complex}. Stacks of studies have already discussed various important aspects of cascading dynamics in complex networks, including modeling the process of failure propagation\cite{motter2002cascade,crucitti2004model,wang2009cascade,buldyrev2010catastrophic,schafer2018dynamically}, vulnerability and robustness of networks \cite{heide2008robustness,wang2008universal,huang2011robustness,zeng2012enhancing}, mitigation and recovery strategies for cascading failures \cite{motter2004cascade,hu2016recovery, smolyak2020mitigation}, and the nature of phase transition induced by attacks\cite{kornbluth2018network,artime2020abrupt}. 

One of the most paradigmatic models depicting propagation mechanism of cascading failure is the Motter-Lai model\cite{motter2002cascade}, which shows the non-local faults diffusion driven by global load redistribution. Non-local cascades have been observed in real systems like blackouts or air-traffic disruptions, all of which can cause global damage to systems\cite{witthaut2015nonlocal}. In particular, cascade size distribution is a basic and crucial characteristic of systemic cascade behaviors\cite{watts2002simple}. From the empirical data, it has been shown that the blackout size distribution approximately follows a power law indicating that the power grid is in a state of self-organized criticality\cite{carreras2016north}. Many cascade models have been established to reproduce power-law distributions of cascade sizes\cite{crucitti2004model, yang2017small,nesti2020emergence}. However, recent studies have suggested that the cascade sizes in complex networks  follow a bimodal distribution with either very small cascades or large ones, which is a typical appearance that usually shows up in the abrupt phase transition\cite{pahwa2014abruptness,spiewak2018study,burkholz2018explicit,artime2020abrupt,kornbluth2021distribution}. The abrupt transition is so sudden and devastating that it is harder to be predicted than a continuous transition. Recently, some researchers have put their attention on the bimodal distribution of cascade sizes\cite{valdez2020cascading,kornbluth2021distribution}. 

Previous research works have revealed that both power-law and bimodal distributions of cascading failures exist in various cascade models. The presence of power-law and bimodal distributions have a dependence on network topology, model parameters, and rules of cascading dynamics\cite{valdez2020cascading,kornbluth2021distribution}. However, current research works focus more on power-law distribution and few research works on bimodal distribution of cascades. These works only show the bimodal distribution phenomenon observed in various cascade models, however, the properties and formation of the bimodal distribution still remain to be explored. Understanding the formation of the bimodal distribution of cascading sizes would help us not only to  predict the damage caused by the cascading dynamics, but also to design methods to mitigate catastrophic cascades. Therefore, it is necessary to get an exhaustive comprehension on the rules behind bimodal cascading sizes and further aid in prediction of the final cascade size.   

In this paper, we investigate in detail the formation of the bimodal distribution of cascades, aiming to distinguish that whether a node belongs to the category in which its breakdown completely disrupts the whole network or leaves it almost intact. To this end, we study the characteristics of nodes among the networks triggering different peaks of the bimodal distribution of cascade sizes after a single initial failure. We find that the final cascade size has close ties with the maximal load of nodes that failed at the first step of the cascade as well as the number of sequential rounds of cascades triggered by the initial failure. Based on this, we propose a hybrid load metric (HLM), in which the load of the initial failed node and the maximal load of First Failures (the failed nodes at the first step of a cascade) are considered, to predict the final cascade size. We then employ AUC, defined by the area under the receiver operating characteristic curve, to measure the accuracy of HLM in identifying a cascade belonging to the left or right peak of the bimodal distribution. By validating on synthetic and real-world networks, we show that HLM outperforms the commonly  used  network centrality metric, betweenness, in predicting final cascade sizes.

\section{Cascading failure model}

\begin{figure*}[h]
    \centering
    \includegraphics[width=1\textwidth]{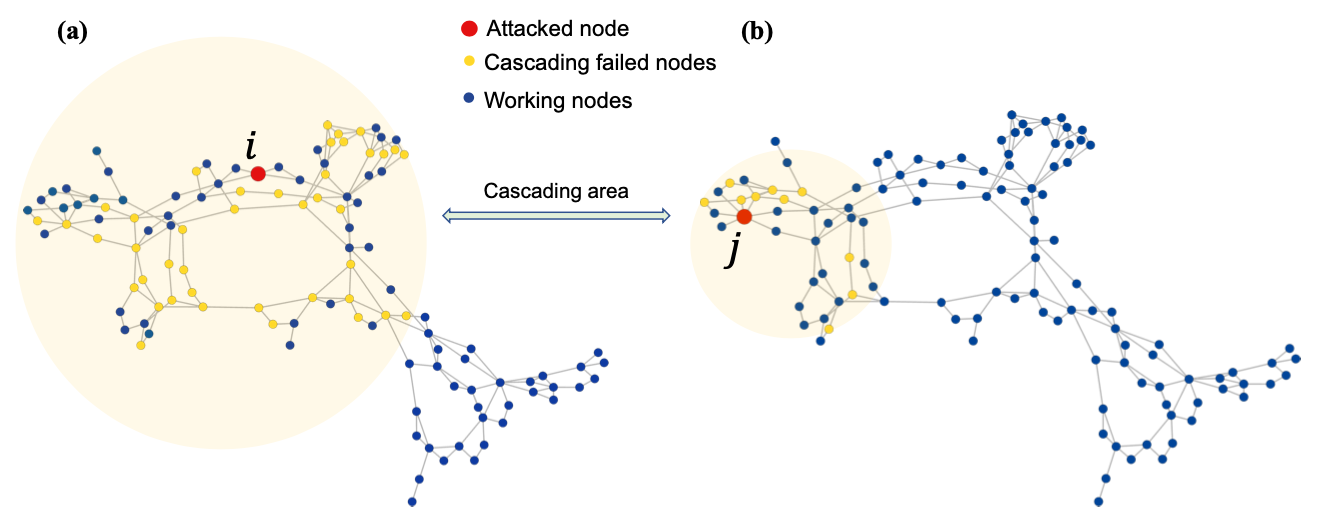}% Here is how to import EPS art
    \caption{\label{shiyi}\textbf{Illustration of differences in cascade sizes triggered by two individual nodes in a power grid}. The network is IEEE 118 Bus $\left(N=118,\langle k \rangle=1.58\right) $, in which nodes represent buses and links represent transmission lines. The initial attacked node is marked in red. The red nodes in \textbf{(a)} and \textbf{(b)} are with a similar load. The yellow nodes are subsequent failures triggered by removing the red node and the links connecting to it. The Motter-Lai model is employed to simulate the cascading dynamics here.}
\end{figure*}

Real networks sometimes undergo undesirable attacks or errors of nodes (edges), which can trigger further failures in other components. There is an established fact that cascading failures can range from only sporadic components failing to catastrophic collapse,  as shown in Fig.~\ref{shiyi}. In IEEE 118 Bus\cite{christie1993power}, the initial failure of node $i$ and node $j$ with similar load trigger completely different cascade sizes. The former causes nearly 40\% of nodes to fail, while the latter only leads to a few nodes being broken. Recent studies have indicated that the distribution of avalanche sizes is bimodal in which some nodes only trigger a limited scope of cascading failures, while others would cause fatal global cascades. In this paper, we are mainly concerned about that how the bimodal distribution of cascades forms and how to distinguish whether a node is capable to bring a widespread network crash that results in a cascade size belonging to the right peak of the bimodal distribution.
%fig1

First of all, we briefly introduce the model of cascading failures that we employed. We consider a simple overload model, which is  proposed by Motter and Lai\cite{motter2002cascade}. In the Motter-Lai model, node $i$ has a certain load $L_i$ given by betweenness, which is the number of shortest paths through this node. $\left(1+\alpha \right)$ times the initial load of node $i$ ($L_i$) is defined as its capacity $C_{i}$ , which is the maximum load  that node $i$ can carry. 
\begin{align}
    C_{i}=\left(1+\alpha \right)L_i ,  \qquad i=1,2,3...N,
\end{align}
where $\alpha$ is a tunable parameter, called the tolerant parameter. $N$ is the number of nodes in the network. In this paper, we fix $\alpha=0.1$, a relatively small value, so that the cascade sizes of nodes are not so large and the cascading results are largely affected by the properties of the initially failed node. In a network, once a node is attacked, all edges connected to this node are also deleted. Therefore, some shortest paths may be diverted, resulting in a change of the load of remaining nodes. If the load of remaining nodes exceed their capacities, these nodes will also be removed. Repeat above process until no node fails in the network, as the end of a cascade. The number of times that this process repeats is donated as $T$, which indicates the number of rounds of cascades triggered by the initial failure. To simplify the description, the first round of cascades triggered by the initial failure is called the first step of a cascade, and so forth. This model is paradigmatic of non-local propagation of failures and widely used in research works to study various faults propagation. 

\section{Results}
\begin{figure*}[h]
    \centering
    \includegraphics[width=1\textwidth]{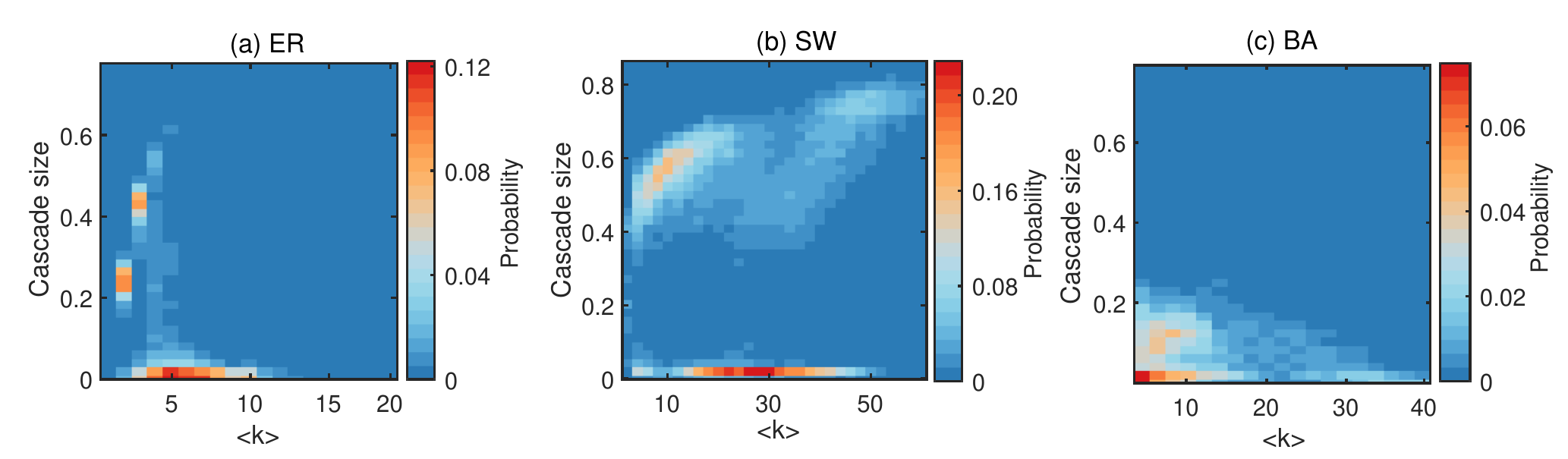}
    \caption{\textbf{The distribution of cascade sizes in synthetic networks with different average degree $\mathbf{\langle k \rangle}$}. \textbf{(a)}, \textbf{(b)} and \textbf{(c)} show the results in ER networks, SW networks and BA networks respectively. The cascade is triggered by a single initial broken node in the network. We get the cascade size of each node and then take binning frequency statistics, which is shown by the heat map. The redder the color is, the higher frequency of such cascade sizes is within the corresponding interval in networks with a specific average degree. Two redder regions longitudinally at a fixed $\langle k \rangle$ indicate that there is a bimodal distribution of cascade sizes in networks with the corresponding $\langle k \rangle$. The Motter-Lai model is employed to simulate the cascading dynamics. The results are computed from the 10 independent networks with a fixed $\langle k \rangle$ and $N=500$.}
    \label{pingjundu}
 \end{figure*}

We first investigate the kinds of networks where the bimodal distribution of cascade sizes can emerge. In particular, we consider a vital property of networks, the average degree $\langle k \rangle$, which tremendously affects the ability of networks to withstand failures. As for information and disease transmission, it is easier for networks with a higher average degree to have wider propagation.\cite{watts2002simple}. However, cascading failure dynamics is different. For example, a network with a higher $\langle k \rangle$ , in which more nodes can share the redistributed load of the failed nodes, is less inclined to trigger a large cascade. Therefore, we first study how the average degree of the network impacts the distribution of cascade sizes. The relative size of a cascade is quantified by the ratio of failed nodes after the initial attack in the network. We stimulate the Motter-Lai model on  Erdős–Rényi (ER) random networks, Small-world (SW) networks and Barabasi-Albert (BA) networks, respectively. In each realization, we attack one node (delete this node and all edges connecting to it) and stimulate the Motter-Lai model to calculate final cascade sizes in the network with a fixed $\langle k \rangle$. The size distribution of cascading failures is recorded, until we have attacked all nodes of networks separately. The results are shown in Fig.~\ref{pingjundu}. The redder the color, the higher frequency of such cascade sizes within the corresponding interval in networks with corresponding $\langle k \rangle$. We can clearly see that there are  bimodal distributions of cascades in all synthetic networks, shown by two redder regions longitudinally at a fixed $\langle k \rangle$. For the ER networks and the BA networks, the bimodal distribution of cascades is more likely to occur in networks with a lower average degree. In networks with too small or too large $\langle k \rangle$, cascade size distributions are unimodal and cascade sizes are mainly at a low level close to 0. This phenomenon is resulted from that a network with too small $\langle k \rangle$ is prone to split the network into some separate connected components that can prevent the propagation of failures and those with too large $\langle k \rangle$  have more links to share the redistributed load of failed nodes\cite{kornbluth2021distribution}. The SW networks, however, exhibit different characteristics that the bimodal distribution always exists, even though the average degree $\langle k \rangle$  is large. This indicates that the small-world property might bring more cascading risks for networks. 

%fig2
 
 %fig3
\begin{figure*}[bt]
    \centering
    \includegraphics[width=1\textwidth]{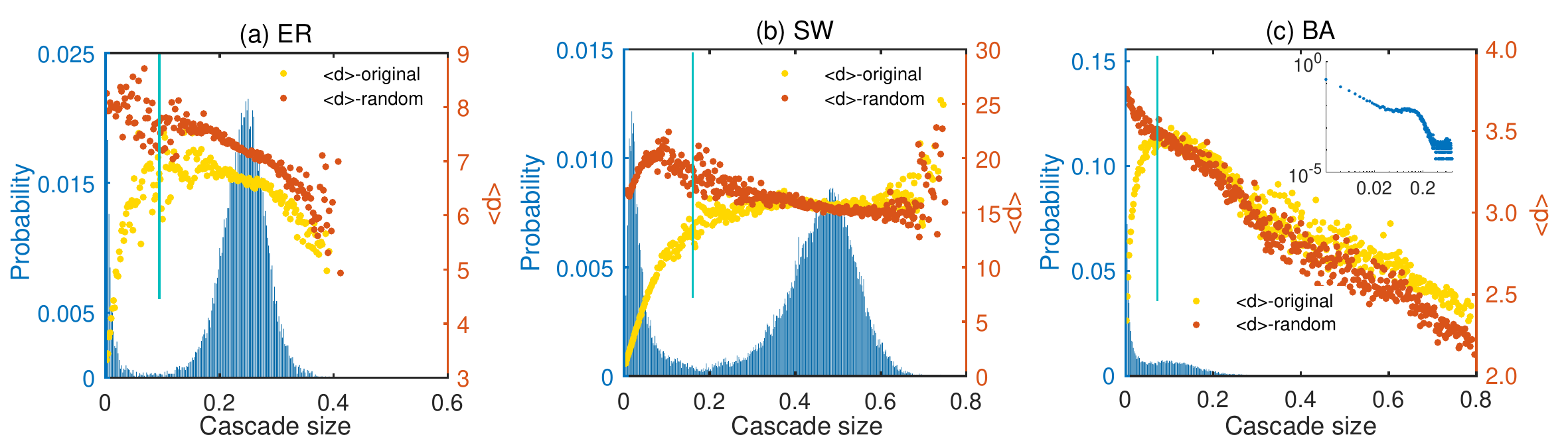}
    \caption{\textbf{The spreading distance $\langle d \rangle$ of failures under different cascade sizes}. The networks are: \textbf{(a)} ER networks $\left(N=500,\langle k \rangle=2\right)$, \textbf{(b)} SW networks $\left(N=500,\langle k \rangle=4\right)$, \textbf{(c)} BA networks $\left(N=500,\langle k \rangle=4\right)$. The spreading distance $\langle d \rangle$ shown by yellow scatters is defined as the average topological distance between the initial attacked node and subsequent failed nodes. Here, subsequent failed nodes (failures) refer to all failures triggered by the initial single failure. And the $\langle d \rangle$-random shown by orange scatters is the result of randomized $\langle d \rangle$ that is obtained by recalculating the mean distance  between randomly selected failed nodes (preserving the failure size) and the initial broken node. The insert in \textbf{(c)} is the distribution of cascade sizes in the double logarithm coordinate. The solid green line indicates the inflection point of $\langle d \rangle$. The above results are computed in 100 independent networks with the same $\langle k \rangle$.}
    \label{shuangfeng}
\end{figure*}

%fig4
\begin{figure*}[tb]
    \centering
    \includegraphics[width=1\textwidth]{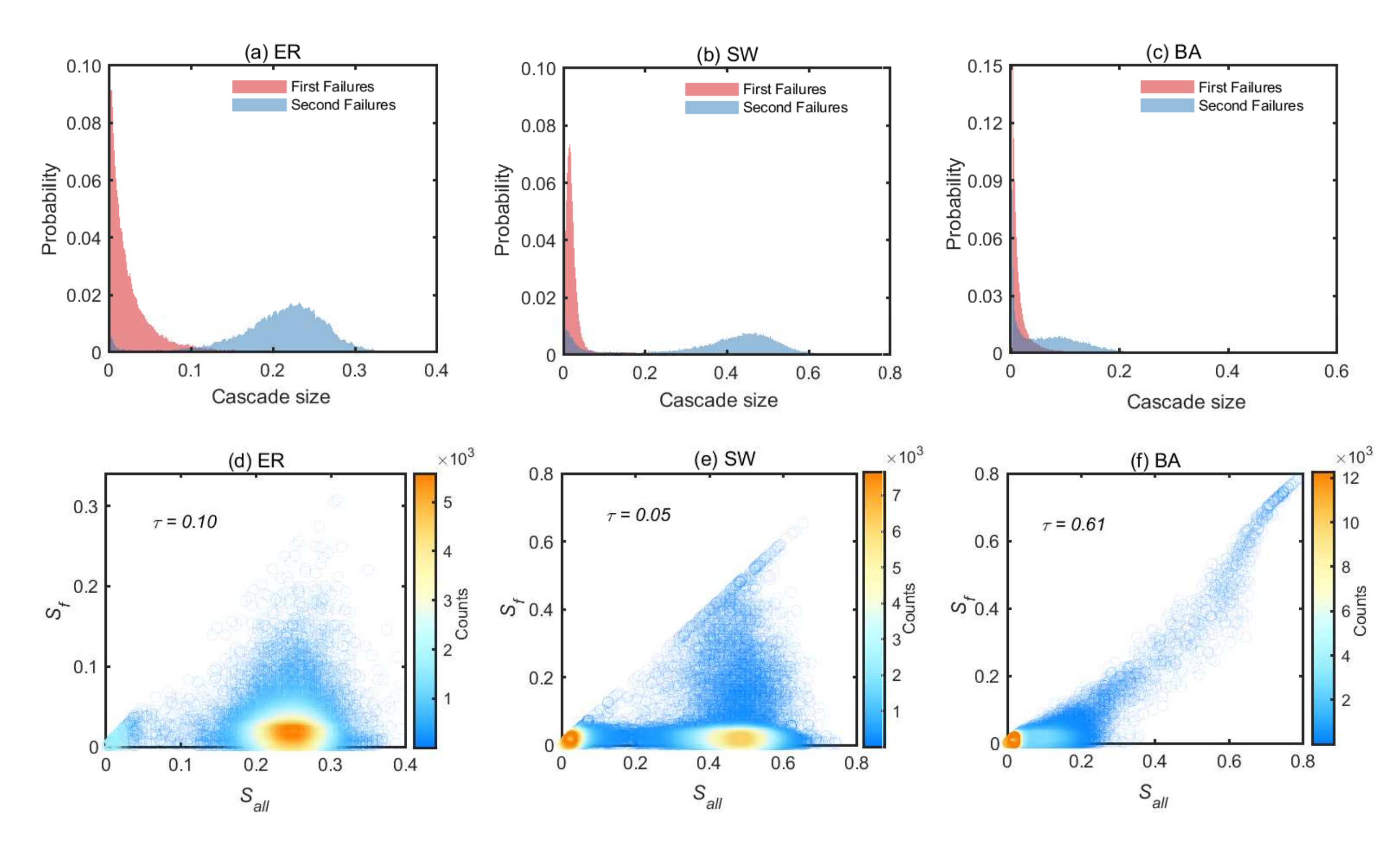}
    \caption{\textbf{The relationship between First Failures and final cascade failures}. Networks are : \textbf{(a, d)}  ER networks  $\left(N=500, \langle k \rangle=2\right)$, \textbf{(b, e)} SW networks $\left(N=500, \langle k \rangle=4\right)$, \textbf{(c, f)} BA networks $\left(N=500, \langle k \rangle=4\right)$. \textit{First Failures} is defined by the failures (failed nodes) at the first step of a cascade and those in the second and later steps is called \textit{Second Failures}. \textbf{(a)}, \textbf{(b)} and \textbf{(c)} show the distribution of First Failures sizes and Second failures sizes in corresponding synthetic networks. \textbf{(d)}, \textbf{(e)} and \textbf{(f)} show the relationship between First Failures sizes ($S_f$) and final cascading failures sizes ($S_{all}$) by the intensity scatter. The color of each point represents the density of nodes within a rectangular area centered on it. The Kendall’s tau rank correlation coefficient $\tau$ is applied to estimate how the First Failures size is correlated to the final cascade size. The above results are computed in 100 independent networks with the same $\langle k \rangle$.}
    \label{yici}
\end{figure*}

The bimodal distribution of cascades has been observed in three different paradigmatic synthetic networks. It is natural to consider what are the differences between the cascades in different peaks of the bimodal distribution. To answer this, we further analyze the relationship between the spreading distance $\langle d \rangle$ of failures and cascade sizes to understand the impact of the bimodal distribution on the propagation range. The spreading distance $\langle d \rangle$ is defined as the average topological distance between the initial attacked node and subsequent failures, which can imply the nonlocality  of failure propagation\cite{witthaut2015nonlocal}. A low $\langle d \rangle$ means that the cascading failures are near the initial breakdown (locality), otherwise are far from the initial one (nonlocality). The bimodal distribution of cascades, in which the left peak separates from the right peak by a gap where cascades rarely occur at these sizes, is shown by a bar chart in different synthetic networks in Fig.~\ref{shuangfeng}. The horizontal axis represents cascade size and the blue bars represent the corresponding probability, which is the proportion of a fixed cascade size among cascade sizes of 50000 nodes from 100 independent networks with the same average degree $\langle k \rangle$. We can see that the bimodality is relatively weak and cascade sizes are also lower in BA networks, which reflects the robustness of BA networks against cascading failures. For spreading distance  $\langle d \rangle$, shown by yellow scatters, it grows as the cascade size increases and then roughly keeps the same with the randomized $\langle d \rangle$ ($\langle d \rangle$-random), which is obtained by recalculating the mean distance  between randomly selected failed nodes (preserving the failure size) and the initial broken node. As the reference of real  $\langle d \rangle$, the randomized $\langle d \rangle$ represents a random counterpart where the failures are cascaded randomly and globally. Obviously, there is an inflection point of spreading distance $\langle d \rangle$ around the gap between two peaks in these networks. To the left of the inflection point, $\langle d \rangle$ increases as the cascade size grows and is significantly lower than the randomized $\langle d \rangle$. However, to the right of the inflection point, the average distance $\langle d \rangle$ is close to the randomized $\langle d \rangle$, which indicates that such cascade sizes in the right peak could cause the same range of damage as the random counterpart. As the results shown above, the cascade sizes in the right peak are nonlocal during the propagation of failures, which can destroy the network globally and randomly. Note that the average distance is slightly lower than the randomized one on the right of the inflection point in ER networks. This implies that the failure propagation in homogeneous networks tends to be more local than those in heterogeneous networks, when compared to a random counterpart.

%fig5 
\begin{figure*}[tb]
    \centering
    \includegraphics[width=1\textwidth]{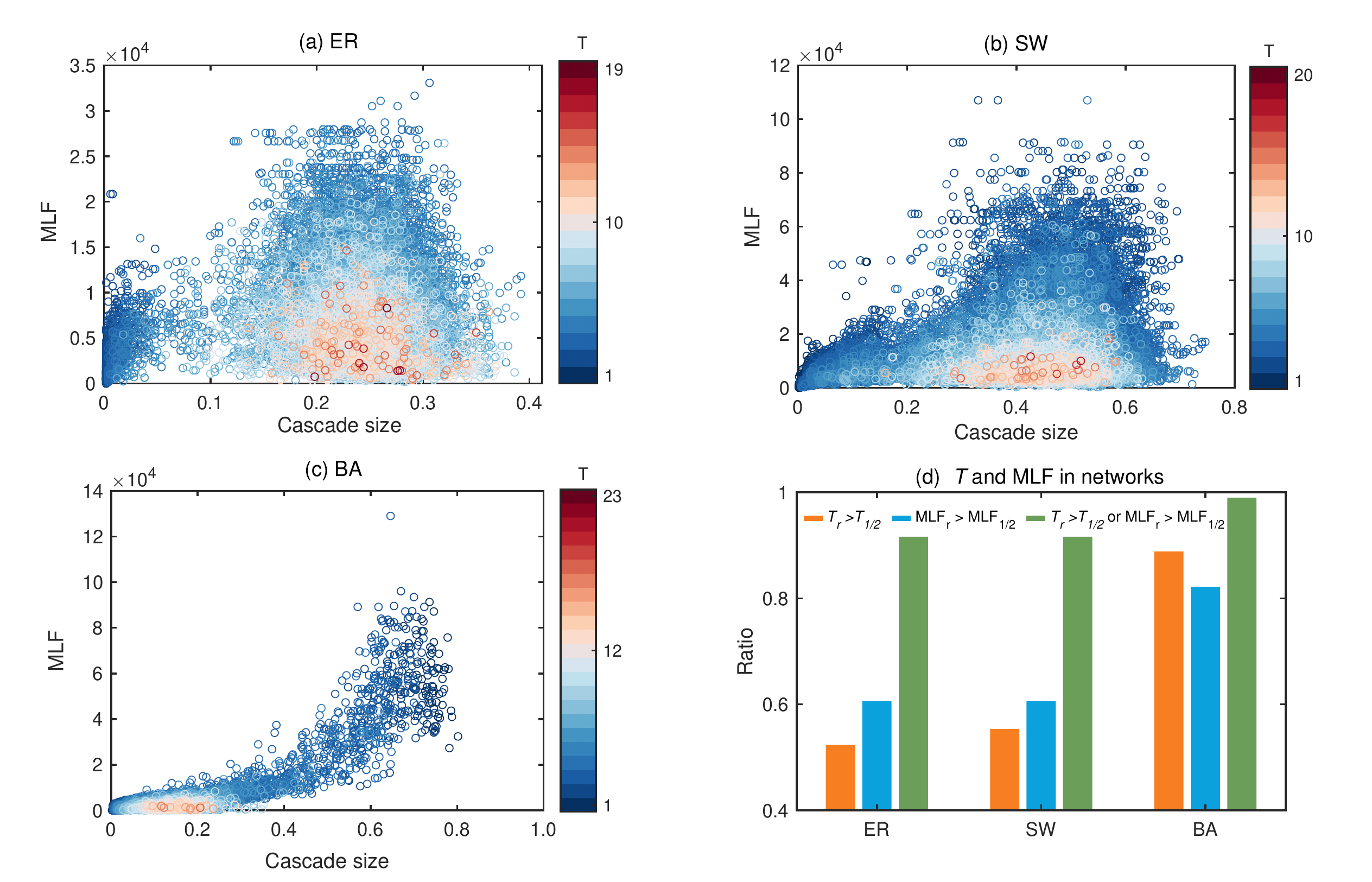}
    \caption{\textbf{The dependence of the cascade size on the maximal load of First Failures (MLF) and cascading rounds $T$ triggered by the initial failure in different synthetic networks}. Networks are : \textbf{(a)} ER networks $\left(N=500, \langle k \rangle=2\right)$, \textbf{(b)} SW networks $\left(N=500, \langle k \rangle=4\right)$, \textbf{(c)} BA networks $\left(N=500, \langle k \rangle=4\right)$. \textbf{(d)} shows the characteristics of MLF and $T$ in these networks. The blue bar is the ratio of nodes in the right peak with higher MLF than the median MLF of the whole data ($\rm{MLF_{r}>MLF_{1/2}}$). The orange bar has a similar meaning that the ratio of nodes in the right peak with higher $T$ than the median $T$ of the whole data. The green bar shows the ratio of nodes with higher MLF or $T$ in the right peak than the corresponding median of the whole data. The above
results are computed in 100 independent networks with the same $\langle k \rangle$.}
    \label{max}
\end{figure*}

The spreading distance $\langle d \rangle$ of failures in the two peaks shows nonlocality characteristics of cascades. The nonlocality of the cascade suggests that the damage is distributed globally, which is hard to predict and recover. Indeed, figuring out the formation of bimodal distribution is the key step in predicting and preventing large-scale cascades. 
Next, we aim to investigate how the bimodal distribution forms. Firstly, we consider intuitively whether the failure size at the first step of a cascade determines the final cascade size. We donate the failures (failed nodes) at the first step of a cascade as \textit{First Failures} and those at the second and later steps are called \textit{Second Failures} here. In Figs.~\ref{yici} (a), (b) and (c), we show the distributions of First Failures sizes and Second Failures sizes in ER networks, SW networks and BA networks respectively. Interestingly, the distribution of First Failures sizes is unimodal and narrow. The result is totally different from the original bimodal distribution, which implies that First Failures cannot represent the final cascading failures. However, the distribution of Second Failures sizes shows weak bimodality, consistent with the original bimodal distribution. In addition, we quantify the dependence of the final cascade sizes on First Failures by the Kendall’s tau rank correlation coefficient\cite{kendall1938new} $\tau$. In Figs.~\ref{yici}(a), (b), we do not 
observe obvious linear relation between final cascade sizes and First Failures sizes, and the Kendall’s tau rank correlation coefficients are also small. In Fig.~\ref{yici}(c), the correlation coefficient is slightly higher. This may be because of high heterogeneity in BA networks, considering the initial failed nodes with high heterogeneity would play decisive roles in their cascading outcomes. To sum up, the final bimodal distribution of cascade sizes is mainly caused by the Second Failures and has no significant correlation with First Failure in ER networks and SW networks. This is significantly different from the propagation of disease and information, where the wider spread probably emerges from more infected agents at the first step of propagation\cite{2014Searching}.

 %fig6
\begin{figure*}[b]
    \centering
    \includegraphics[width=1\textwidth]{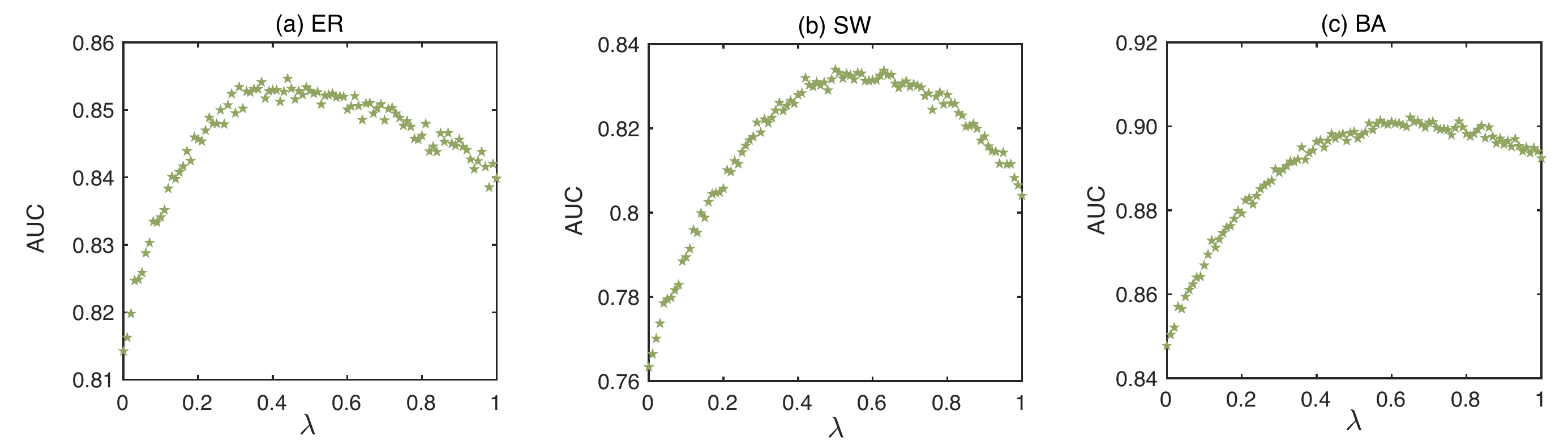}
    \caption{\textbf{The AUC value of HLM under different parameter $\lambda$}. The networks are:  \textbf{(a)} ER networks $\left(N=500, \langle k \rangle=2\right)$,  \textbf{(b)} SW networks $\left(N=500, \langle k \rangle=4\right)$,  \textbf{(c)} BA networks $\left(N=500, \langle k \rangle=4\right)$. The results are obtained by averaging 10 independent realizations.}
    \label{lam}
\end{figure*}

\begin{table*}[htbp]
    \centering 
    \caption{\textbf{Structural properties and AUC value of the different metrics in synthetic and real networks}. Structural properties include network size $(N)$ and average degree ($\langle k \rangle$). Metrics include the load of the initial attacked node $(L)$, size of First Failures ($S_1$), the spreading distance ($d_1$) at the first step that is defined by the average distance between the initial attacked node and those of First Failures, the maximal load of First Failures (MLF) and HLM with an optimal $\lambda$ achieving the largest AUC value ($\rm{HLM^*}$).}
    % \begin{ruledtabular}
    \begin{tabular}{cccccccc}
    % {p{2cm}<{\raggedright}p{1cm}<{\centering}p{1cm}<{\centering}p{1.5cm}<{\centering}p{1.5cm}<{\centering}p{1.5cm}<{\centering}p{1.5cm}<{\centering}p{1.5cm}<{\centering}}
    \hline\hline
    \textbf{Networks} & $\mathbf{N}$ & $\mathbf{\langle k \rangle}$ & $\mathbf{\langle d_1 \rangle}$ & $\mathbf{S_f}$  & $\mathbf{L}$ & \textbf{MLF} & {{$\mathbf\rm{HLM^*}$}}\\
    \hline
    BA & 500 & 4 & 0.8303 & 0.8375 & 0.8483 & 0.8924 & 0.9022\\
    ER & 500 & 2 & 0.7891 & 0.7833 & 0.8157 & 0.8405 & 0.8546\\
    SW & 500 & 4 & 0.5747 & 0.5597 & 0.7646 & 0.8062 & 0.8339\\
    Bus 1494 & 1494 & 2.89 & 0.7763 & 0.7967 & 0.8145& 0.8453 & 0.8459 \\
    Euro-road & 1039 & 2.51 & 0.5000 & 0.7185 & 0.6816 & 0.7949 & 0.7949 \\
    Bus 494  & 494 & 2.37 & 0.7800 & 0.7752 & 0.8134 & 0.8750 & 0.8785 \\
    Econ-poli & 2343 & 2.28 & 0.7772 & 0.8053 & 0.8455 & 0.9038 & 0.9039 \\
    Bio-yeast & 1458 & 2.67 & 0.6798 & 0.8140 & 0.8263 & 0.8263 & 0.8821\\
    \hline\hline
    \end{tabular}
    \label{biaoge}
    % \end{ruledtabular}
\end{table*}

%fig7
\begin{figure*}[h]
    \centering
    \includegraphics[width=0.98\textwidth]{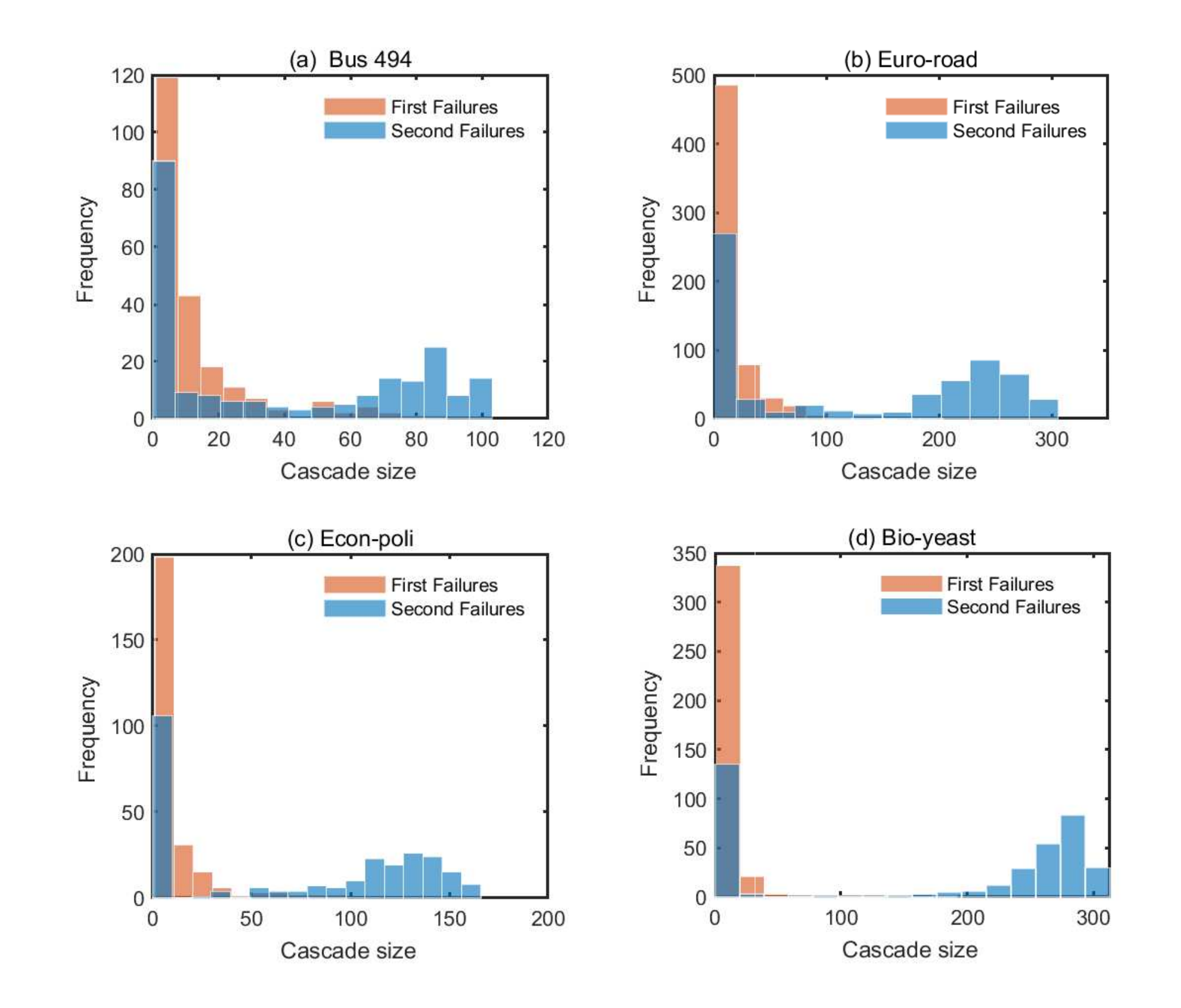}
    \caption{\textbf{The distribution of First Failures sizes and Second Failures sizes in real-world networks.} The networks are:\textbf{(a)} Bus 494 $\left(N=494, \langle k \rangle=2.37\right)$, \textbf{(b)} Euro-road $\left(N=1039, \langle k \rangle=2.51\right)$,  \textbf{(c)} Econ-poli $\left(N=2343, \langle k \rangle=2.28\right)$, \textbf{(d)} Bio-yeast $\left(N=1458, \langle k \rangle=2.67\right)$.}
    \label{rel}
 \end{figure*}

Based above analysis, the size of First Failures does not determine the final cascading size to a large extent, which implies that it cannot be used to predict the final cascade size. Now, we turn our attention to the characteristics of First Failures to see  whether they influence the final cascade size. The load of nodes plays an essential role, which not only determines the redistributed flow, but also the capacity of  nodes in the cascading dynamics. Therefore, we further explore the dependence of the cascade size on the maximal load of First Failures (MLF) in different artificial networks. The number of rounds of cascades $T$ triggered by the initial failure as a vital variable of cascading dynamics is also considered here. The results are shown in Fig.~\ref{max}. Each cascade is triggered by a single node. MLF and $T$ are recorded after the cascade ends. Therefore, every node will be with a cascade size, an MLF value and a $T$ value. Beforehand, we have observed that the distribution of cascade sizes is bimodal in these networks in Fig.~\ref{shuangfeng}. The boundary between the left peak and right peak is defined by the cascade sizes with the lowest probability in the gap of two peaks here. In Figs.~\ref{max} (a), (b) and (c), it is obvious that the large cascade sizes distributed in the right peak are associated with higher MLF and $T$ than those in the left peak. To characterize this phenomenon more precisely, we further measure the values of MLF and $T$ of the cascades in the right peak compared with the whole data. We first calculate the median MLF value and $T$ value of all nodes in networks ($\rm{MLF_{1/2}}$, $T_{1/2}$). Then, we calculate the ratio of nodes in the right peak with higher MLF values than $\rm{MLF_{1/2}}$. The higher such ratio is, the more nodes would be with higher MLF values in the right peak. Similarly, the ratio of nodes in the right peak with higher $T$ values than $T_{1/2}$ is recorded. Moreover, the union of them that the percentage of nodes in the right peak with higher MLF values or $T$ values than their
corresponding median of the whole data is also considered. The results are shown in Fig.~\ref{max} (d). We can see that the ratios are over 0.5 and the union of them is more than 0.9 in these networks, which indicates that most large cascades in the right peak tend to be with either higher MLF values or higher $T$ values. A higher MLF value means that the nodes with high load failed at the first step of the cascade and a higher $T$ means that more rounds of cascades triggered by the initial failure. In other words, most large cascades in the right peak is resulted from either explosive breakdown in only a few steps or accumulation of multiple rounds of cascades triggered by the initial failure. 

The higher MLF and $T$ in the right peak where the larger cascade sizes are distributed, can provide us a new view of predicting the sizes of cascading failures. The  MLF value only contains the information from the first step of the cascade, although $T$ is derived from the entire process of cascading dynamics. In the previous studies, the load of the initial attacked node is always considered to predict the final cascade size. However, there are some scenarios ignored, in which the final damage is not strongly dependent on the initial failure but closely related to the detailed cascade process. For instance, a node with a low load failed might trigger the failure of a core node, which would cause a large cascade. Accordingly, we propose a hybrid load metric (HLM), combining the load of initial failure and MLF with a tunable parameter, to predict the final cascade size after an initial attack. HLM value of node $i$ is defined as

\begin{eqnarray}\label{eq2}
    \centering
    {\rm{HLM}}_i=\lambda{L}_i+\left(1-\lambda\right){\rm{MLF}_i},
\end{eqnarray}
where $L_{i} $ is the load of the initial attacked node $i$. $\rm{{MLF}_i}$ is the maximal load of First Failures triggered by the initial attacked node $i$. $\lambda $ is a tunable parameter controlling their weight. Actually, when $\lambda=0$, HLM is equivalent to MLF. When $\lambda=1$, HLM returns to the commonly used centrality index, namely the betweenness (load) of the initial failed node. To validate the effectiveness of HLM, we employ Area Under Curve (AUC)\cite{1982The} to evaluate its effectiveness for predicting the final cascade size. We randomly choose a node in the left and right peaks respectively and then compare their HLM values at each time. In $n$ independent comparisons, if there are $n^{\prime}$ times the node in the right peak having a higher HLM and $n^{\prime\prime}$ times they have the same HLM value, the AUC is expressed as
\begin{eqnarray}
    \centering
    {\rm{AUC}}=\frac{n^{\prime}+0.5n^{\prime\prime}}{n}.
\end{eqnarray}
The AUC value is about 0.5 for the random prediction. The AUC value can be interpreted as the probability that a randomly
chosen node in the right peak is given a higher HLM value than that in the left peak. A higher AUC value implies that the metric has a better performance on prediction or classification. We first investigate how the parameter $\lambda $ affects the AUC value in synthetic networks, which is shown in Fig.~\ref{lam}. We can see that there is an optimal $\lambda $ achieving the largest AUC in three synthetic networks and it is apparently higher than the AUC value when $\lambda=0$ or 1. In addition, AUC with $\lambda=0$ is higher than that with $\lambda=1$, which means that MLF has a better performance than $L$ in predicting the final cascade size. Hence, the MLF value can better symbolize the final cascade size compared to the load of initial broken node $L$, which is commonly used. The detailed results are reported in Table.~\ref{biaoge}. For comparison, we also test other general metrics including the load of the initial attacked node $L$, the size of First Failures $S_f$, and the spreading distance $\langle d_1 \rangle$ at the first step of cascades which is the average distance between the initial attacked node and those of First Failures. We can see that  MLF and  HLM apparently outperform other metrics, and HLM is the best one. In fact, HLM combining the load of initial failure and information of subsequent failed nodes, makes up for the deficiency of only focusing on the load of the initial failed node by taking the cascading dynamics at the early stage of cascades into consideration.

To further support the effectiveness of our method, we apply it to real networks including power grids (Bus 1494, Bus 494)\cite{nr-aaai15}, a road network (Euro-road)\cite{10.1145/2487788.2488173}, an economic network (Econ-poli)\cite{nr-aaai15} and a biological network (Bio-yeast) \cite{jeong2001lethality}. Their structural properties are shown in Table.~\ref{biaoge}. Likewise, there is a bimodal distribution of cascade sizes in each of these real networks, part of which are shown in Fig.~\ref{rel}. First failures and Second Failures also obey the same features as in synthetic networks that the distribution of First Failures sizes is unimodal and the distribution of Second Failures sizes shows weak bimodality. The AUC of the metrics in real networks is displayed in Table.~\ref{biaoge}. MLF and HLM also outperform other metrics, which is consistent with those observed in synthetic networks. 

\section{Discussion}  

 The cascade size is a big concern when researchers discuss cascading failures. Numerous research works have focused on the power-law distribution of cascade sizes, which implies that the system is driven by a self-organized state. However, recent studies suggest that the distribution of cascade sizes is bimodal in which the cascade size triggered by one node is either very small or very large. Indeed, both power-law and bimodal distributions of cascading failures exist in complex networks, which depend on network topology, model parameters and rules of cascading dynamics. Previous studies have intensively discussed the power-law cascade size distribution. However, the bimodal distribution is rarely involved. Therefore, it is necessary to study the rules and characteristics  behind the bimodal distribution, which can help to understand deeply cascading dynamics and aid prediction of the final cascade size. 

 In this article, we focus on the formation and prediction of bimodal distribution of cascade sizes. What we do first is to investigate the features of nodes in different peaks of the bimodal distribution. The results show that the large cascade sizes distributed in the right peak are resulted from either the failure of higher-load nodes at the first step or triggering multiple rounds of cascades after the initial failure. In other words, large-scale cascading failures are the results of either an explosive breakdown in only a few steps or cumulative multiple rounds of cascades triggered by an initial attack. In this context, we propose MLF (the maximal load of First Failures) and a hybrid load metric (HLM) combining the load of the initial attacked node and MLF with a tunable parameter $\lambda$. Then, employ them to distinguish whether the breakdown of a node completely disrupts the whole network or leaves it almost intact, which respectively correspond two peaks of the bimodal distribution. By comparing the AUC value, both MLF and HLM show better performance compared to common indicators, such as the load of the initial attacked node and the size of First Failures. Moreover, real-world networks also show similar results.

The bimodal distribution of cascade sizes is an interesting phenomenon that deserves further effort to investigate the mechanism behind it. Our work reveals that the statistical properties of bimodal cascade size distribution and propose an effective metric HLM to estimate the size of cascading failures triggered by a single node. Despite its effectiveness, there are several limitations of it. For example, HLM can achieve only a rough distinction of cascades between two peaks of bimodal distribution, not a highly accurate prediction. More accurate prediction methods are asked for further exploration. Moreover, the information of the first step in cascade progress is also involved in HLM, which means we cannot predict by HLM before the cascades occur. In fact, the prediction of the failure spreading is rarely studied, because cascades are the results of global interactions among nodes which would make the prediction harder. Our object is to predict the final cascade at the early stage of cascades. Finally, our research only considers a simple case where the cascades are triggered by a single node, and does not involve the cascades due to massive initial failures, which is also an important research direction of cascading dynamics asked for further investigation.

\clearpage

\bibliography{biomodal}

\section*{Acknowledgements}
This work was supported by National Natural Science Foundation of China (71731002).

\section*{DATA AVAILABILITY}
The data that support the findings of this study are available from the corresponding author upon reasonable request.

\end{document}